\documentstyle[12pt,epsf]{article}

\def\be{\begin{equation}}
\def\ee{\end{equation}}
\def\bea{\begin{eqnarray}}
\def\eea{\end{eqnarray}}
\def\nn{\nonumber}

\newcommand{\Section}[1]{\section{#1}\setcounter{equation}{0}}

\begin{document}

\begin{flushright}
hep-th/0108095 \\
TIFR/TH/01-29
\end{flushright}

\pagestyle{plain} 

\def\e{{\rm e}}
\def\cs{\frac{1}{(2\pi\alpha')^2}}
\def\CV{{\cal{V}}}
\def\haf{{\frac{1}{2}}}
\def\tr{{\rm Tr}}
\def\"{\prime\prime}
\def\p{\partial}
\def\tphi{\tilde{\phi}}
\def\ttheta{\tilde{\theta}}
\def\hj{\hat j}
\def\hn{\hat n}
\def\bz{\bar{z}}
\def\zk{{\bf{Z}}_k}
\def\h1{\hspace{1cm}}
\def\goes{\rightarrow}
\def\goal{\alpha'\rightarrow 0}

\vspace{3cm}

\begin{center}
{\Large {\bf Effective Action for D-branes on SU(2)/U(1) Gauged WZW
Model}}
\end{center}

\vspace{.15cm}

\begin{center}
Shahrokh Parvizi
\footnote{On leave from {\it ``Institute for Studies in
Theoretical Physics and Mathematics"}, Tehran, Iran.}

\vspace{.3 cm}

{\it Department of Theoretical Physics,\\
Tata Institute of Fundamental Research,\\
Homi Bhabha Road, Mumbai 400 005, INDIA.}\\

\vskip .1 in 
%

{\sl parvizi@theory.tifr.res.in}
\vspace{1cm}
\end{center} 
\vskip .02 in 
\begin{abstract}
Dynamics of D-branes on $SU(2)/U(1)$ gauged WZW model are investigated. We
find the effective action for infinite $k$, where $k$ is the level of WZW
model. We also consider finite $k$ correction to the effective action
which is compatible with Fedosov's deformation quantization of the
background.

\end{abstract}

\newpage
\Section{Introduction}

Initially, D-branes in string theory were defined as hyper-surfaces on
which open strings can end \cite{polchinski}.  More precisely, the branes
can be considered as boundary states in the close string picture, subject
to some gluing condition \cite{green}. The earlier works of Ishibashi
\cite{ishibashi} and Cardy \cite{cardy} for RCFT, make it possible to
construct the boundary states for those backgrounds on which the
representation of string states are known.

On the other hand, it is known that there is a non-commutative structure
on the branes in a flat background \cite{NC,SW}. This non-commutativity
can be formulated by a $*$ product known as Moyal product with a constant
non-commutativity parameter $\theta$. Again on a curved manifold, the
Moyal product is ill-defined and the $\theta$, effectively, is
non-constant.

So the important task is, firstly, finding Cardy's solutions for a given
background, secondly, looking for an effective action for which a
consistent non-commutative structure is expected. For the flat space, the
result is a non-commutative gauge theory (or a matrix model for D0-branes)
with Moyal brackets. But, for a general background, it is still an open
problem.

To study this subject, important special classes are group manifolds and
quotient spaces. The D-branes on group manifolds were introduced in
\cite{dgroup} and their effective action and non-commutative structure
were studied in \cite{ARS}. Recently, the quotient space, $SU(2)/U(1)$ in
the frame of gauged WZW models has been studied in \cite{mald}. This space
topologically is a disk with radius $\sqrt{k}$, where $k$ is the level of
WZW model. Several D-branes are introduced on this disk and it is the
purpose of the present paper to study the dynamics of these branes.

This paper is organized as follows. In the second section, we review the
introduction of the space and its branes. In section 3, a D2-brane is
studied by DBI action. We find the mass of D2-brane and the open string
spectrum as fluctuating modes on the brane. In the fourth section, we
compute the effective action from boundary conformal field theory. The
result is a non-commutative $U(1)$ gauge theory. This result is valid only
for $k \goes \infty$, but in section 5, we extend the effective action for
finite $k$ corrections and show that the result is some non-commutative
theory with a non-constant $\theta$. We show that this non-commutativity
is consistent with the known Fedosov's $\star$ product on the disk. The
Fedosov's formalism is a deformation quantization on a curved space which
gives a natural definition for the non-commutativity on the curved space.

\Section{D-Branes on $SU(2)/U(1)$ WZW}

The $SU_k(2)$ WZW model can be considered as a sigma model on $S^3$ with
an antisymmetric B-field. Firstly, consider the $S^3$ metric as follows,

\bea\label{1.10}
ds^2= d\theta^2+ \sin^2\theta d\phi^2 + \cos^2\theta d\tilde{\phi}^2
\eea
for which one can use the Euler's angels parameterization of a group
element, $g \in SU(2)$, 
\bea\label{1.15}
g=\e^{i\chi\sigma^3/2}
\e^{i\tilde{\theta}\sigma^1/2} \e^{i\varphi\sigma^3/2}
\eea 
where,
\bea
\chi= \tilde{\phi}+\phi, \h1 \varphi=\tilde{\phi}-\phi, \h1
\tilde{\theta}=2\theta.
\eea
The anti-symmetric B-field in the Wess-Zumino term of WZW model is as
follows,
\bea \label{1.20}
B= sin^2\theta\; d\phi \wedge d\tphi
\eea
On this background, the WZW action will be,
\bea \label{1.25}
S_{SU(2)}&=& 
k\int d^2z \lgroup 
 \p\theta\bar{\p}\theta+ \sin^2\theta \p\phi\bar{\p}\phi + \cos^2\theta
\p\tilde{\phi}\bar{\p}\tilde{\phi}+\sin^2\theta (\p\phi\bar{\p}\tphi-
\bar{\p}\phi\p\tphi) \rgroup,\\
&=& 
k\int d^2z \lgroup
 \p\theta\bar{\p}\theta+ \tan^2\theta \p\phi\bar{\p}\phi + \cos^2\theta
(\p\tphi+\tan^2\theta\p\phi)(\bar{\p}\tphi-\tan^2\theta\bar{\p}\phi)
 \rgroup.
\eea

Now we consider the gauged model, $SU(2)/U(1)$, by gauging a $U(1)$
corresponding to shifting $\tphi$. It needs introducing of gauge
fields in the action as,

\bea \label{1.30}
S = k\int d^2z \lgroup
 \p\theta\bar{\p}\theta+ \tan^2\theta \p\phi\bar{\p}\phi + \cos^2\theta
(\p\tphi+\tan^2\theta\p\phi+A_z)
(\bar{\p}\tphi-\tan^2\theta\bar{\p}\phi+A_{\bz})\rgroup,
\eea
Then by integrating out $A$ fields, one finds a sigma model with the
following metric and non-constant dilaton \cite{mald}:
\bea \label{1.35} 
ds^2 &=& \frac{k}{1-r^2}(dr^2 + r^2 d\phi^2)\\ \label{1.36}
g_s(r)&\equiv& \e^{\Phi}= g_s(0)(1-r^2)^{-\haf}.
\eea
where $r=\sin \theta$. This is a disk with boundary at $r=1$. Although, it
seems that the target space has $U(1)$ symmetry corresponding to the shift
of $\phi$, actually it is broken to a $\zk$ symmetry. This can be
seen explicitly by finding the divergence of the current of $\phi$ shift,
$\p_\alpha j^\alpha \sim k F_{z\bz}$ where $F$ is the field strength of
$A$ field. Now a $\zk$ subgroup of $U(1)$ is non-anomalous and is a
symmetry of the model.

On the other hands, from the GKO construction, the  $SU_k(2)/U_k(1)$ is
known as {\it parafermion} model. This model has $\zk\times\overline{\bf
Z}_k$ symmetry. A representation in this model can be shown by $(j,n)$
where $j=1, \haf, 1, ..., \frac{k}{2}$ is  
the $SU(2)$ spin and $n= -k+1, -k+2, ..., k$. We need also the
identification $2j + n = 0 \;\;(mod \; 2)$. 

In the parafermion language, the D-branes are introduced as boundary
states. Consider $(J \pm \bar{J})|Boundary>=0$ as gluing conditions with 
corresponding solutions as $A$-states and $B$-states, respectively. The
$A$-solutions can be constructed as Ishibashi states with identical left
and right sectors,
\bea \label{1.40}
|A;jn\gg \; = \sum_{States} |j,n> \otimes \widetilde{|j,n>}
\eea
Then the following linear combination gives Cardy states with
correct modular transformations \cite{cardy},
\bea \label{1.45}
|A;\hj\hn>_C = \sum_{j,n\in PF}
\frac{S_{\hj\hn}^{PF\;jn}}{\sqrt{S_{00}^{PF\;jn}}} |A;jn\gg
\eea
where $S^{PF}$ is the modular transformation matrix of parafermionic
theory and is introduced as follows \cite{mald},
\bea \label{1.50}
S_{\hj\hn}^{PF\;jn}=2\sqrt{\frac{1}{k(k+2)}} \e^{\frac{i\pi n \hn}{k}}
\sin\left( \frac{\pi(2\hj+1)(2j+1)}{k+2}\right).
\eea

These A-branes have geometric interpretations as D0-branes on some special
points on the disk circumstance and also D1-branes as lines between these
special points. Also it is possible to find the B-branes trough T-duality
transformation on the disk.\footnote{This T-duality interchanges the
role of circumstance and center of the disk with the map,
$r'=\sqrt{1-r^2}$, see \cite{mald}.} The results are D2-branes co-centered
with the disk
and D0-branes near the center of the disk.\footnote{Besides these branes,
there are other branes on the disk, but for our purpose, we consider only
the above mentioned B-branes (see \cite{mald}).} 

\Section{DBI Action for D2-Brane with Flux Stabilization}

In $SU(2)$ model, D2-branes are two-spheres and for $j \ll k$, they can be
considered as bound states of $(2j+1)$ D0-branes. These D2-branes go to
flattened spheres or equivalently, two sided disks in $SU(2)/U(1)$ theory.
To support this idea, we follow the procedure of \cite{mald} and
\cite{bachas} to find the open string spectrum on this D2-brane.

Consider a D2-brane as a disk with radius $r_m$ centered at $r=0$. This
brane can be stabilized by an $F$ field with fixed flux equal to the
number D0-branes on it. So for a two sided D2-brane, we
have,\footnote{Take units with $\alpha'=1$.}
\bea \label{2.10}
\frac{N}{k}=\frac{(2\hj+1)}{k}=2\frac{1}{2\pi}
\int_{D2}\frac{F}{k}=\frac{1}{\pi}\int_0^{r_m}dr f(r),
\eea
in which $N=(2\hj +1)$ is the number of D0-branes near the center
and $f(r):= 2\pi F_{r\phi}/k$.

Now, we can minimize the DBI action 
for the above D2-brane in the  given background
(\ref{1.35}) and (\ref{1.36}).
Consider $\lambda$ as a Lagrange multiplier to impose the fixed flux
condition (\ref{2.10}), one can minimize the following quantity:
\bea \label{2.15}
\frac{k}{2\pi g_s(0)}\int d\phi dr \sqrt{1-r^2}\sqrt{r^2/(1-r^2)^2+f^2} -
k\lambda(\frac{1}{\pi}\int_0^{r_m} f(r') dr' -\frac{N}{k}),
\eea  
and the corresponding $f$ will be \cite{mald},
\bea \label{2.20}
f_m=\frac{r}{(1-r^2)}\frac{\sqrt{1-r_m^2}}{\sqrt{r_m^2-r^2}}.
\eea
This field strength can be derived from the following gauge field,
\bea \label{1.22}
A_{\phi}= \frac{k}{2\pi} \tan^{-1}\frac{\sqrt{r_m^2-r^2}}{\sqrt{1-r_m^2}},
\eea
and $r_m$ can be found in terms of the flux \cite{mald},
\bea \label{2.25}
\frac{\pi(2\hj+1)}{k}=\int_0^{r_m} dr f(r) =
\tan^{-1}\frac{r_m}{\sqrt{1-r_m^2}},
\eea
or 
\bea \label{2.30}
r_m = \sin\left(\frac{\pi(2\hj+1)}{k}\right).
\eea

In \cite{bachas}, the mass of D2-brane in $SU(2)$ background is computed
as the minimum of the DBI action. Also by considering small fluctuations
around the classical solution, the open string spectrum has been found
\cite{bachas}. 
Now, we follow \cite{bachas} to find the D-brane mass and open 
string spectrum for $SU(2)/U(1)$ case. Firstly, 
put back $f_m$ in the DBI action, we will find the mass of D2-brane as
follows,
\bea \label{2.35}
M_{D2}=S|_{f_m}=\frac{k}{g_s(0)}r_m=\frac{k}{g_s} 
\sin\left(\frac{\pi(2\hj+1)}{k}\right).
\eea
This mass is comparable with that derived from the overlap of D-brane
state with $|A;j=0,n=0\gg$ state, for large $k$ (see eq. (3.8) in
\cite{mald}).

To find open string spectrum on this D2-brane, we turn on  
small fluctuations around the classical solution as follows,
\bea \label{2.40}
\tilde{A_{\phi}}&=& \frac{k}{2\pi}
\tan^{-1}\frac{\sqrt{r_m^2-r^2}}{\sqrt{1-r_m^2}}+ \frac{k}{2\pi}
a_{\phi}\\ 
\tilde{A_r}&=&  \frac{k}{2\pi} a_r, \;\;\; A_t =0 = \tilde{A}_t
\eea
In terms of these fields, we can write $G+2\pi F$ matrix as,
\bea \label{2.42}
G+2\pi F = k \left(\matrix{\frac{-1}{k} & \p_t a_r & \p_t a_{\phi} \cr
             -\p_t a_r & \frac{1}{1-r^2} & \p_r
\tan^{-1}\frac{\sqrt{r_m^2-r^2}}{\sqrt{1-r_m^2}}+f_{r\phi} \cr
-\p_t a_{\phi} &
-\p_r\tan^{-1}\frac{\sqrt{r_m^2-r^2}}{\sqrt{1-r_m^2}}-f_{r\phi} & 
\frac{r^2}{1-r^2} }\right)
\eea
where $f_{r\phi}:= \p_r a_{\phi} - \p_{\phi} a_r$. 
Inserting these into DBI action and expanding up to second order, we find
the following equation of motion for fluctuating fields,
\bea \label{2.45}
\ddot{f}_{r\phi}= \frac{1}{k}\left(
\p_r\left( \frac{r(1-r^2)}{\sqrt{r_m^2-r^2}} 
\p_r\frac{(r_m^2-r^2)^{3/2}}{r(1-r^2)}\right) +
\frac{r_m^2-r^2}{r^2} \p_{\phi}^2 \right) f_{r\phi}
\eea
For $r_m \goes 1$ and with $r = \sin \theta$, it will be,
\bea \label{2.50}
\ddot{\tilde{f}}= \frac{1}{k}\left(
\frac{1}{\sin \theta}\p_{\theta}\left( \sin \theta
\p_{\theta}\right) +
\frac{1}{\sin^2\theta} \p_{\phi}^2 - \p_{\phi}^2 \right) \tilde{f}
\eea
where $\tilde{f}:= f_{r\phi} \cot \theta$. The above equation can be
written in terms of the
Laplace's operator of $S^2$ and $J_3^2$, the third component of angular
momentum operator, {\it i.e.},
\bea\label{2.55}
\ddot{\tilde{f}}=- \frac{1}{k}\left(
\Box_{S^2}-J_3^2 \right) \tilde{f}
\eea
The eigenvalues are well-known,
\bea \label{2.60}
M^2 = \frac{j(j+1)}{k}-\frac{m^2}{k}.
\eea
This is in agreement with conformal dimensions of the GKO
construction of $SU(2)/U(1)$, for large $k$.

\Section{Effective Action from Boundary CFT}

In this section, we use boundary CFT formalism to find the effective
action of D0-branes. Formally, it is possible to find the effective action
term by term from the $n$-point functions of vertex operators on the
brane. For definiteness, we consider $:J^\mu V[{\rm A}_\mu]:(x_1)$ as
boundary field in which $J^\mu$ is the current and $V[{\rm A}_\mu] =
\sum_{\alpha}a_\mu^{\alpha}V[\alpha]$ is a lowest weight vertex operator
in some suitable basis. Now, the cubic interaction in the effective action
will corresponds to the following 3-point function,\footnote{$\mu, \nu, 
\rho =
1, 2, 3$ for $SU(2)$ case. For $SU(2)/U(1)$ branes, with large $k$ limit,
indices are $a, b, c= 1, 2$, see bellow.} 
\bea \label{3.10}
<:J^\mu V[{\rm A}_\mu]:(x_1) :J^\nu V[{\rm A}_\nu]:(x_2) :J^\rho V[{\rm
A}_\rho]:(x_3)>
\eea
To calculate the above 3-point function, three kinds of OPE are needed.
Firstly, the OPE of currents $J$'s or currents algebra. Secondly, the OPE
of currents and vertex operators which is the definition of current
algebra primaries, and finally, the OPE of vertex operators which
corresponds to the fusion algebra.  

Our case, $SU_k(2)/U_k(1)$ or parafermionic theory, 
has $\zk\times\overline{\bf Z}_k$
symmetry for left and right handed sectors. To realized it, it is common to
introduce currents \cite{zamolodchikov}, 
$\psi_l(z)$ and $\bar{\psi}_l(\bz)$ (for $l=1, 2, ..., k$), with $(l,0)$
and $(0,l)$ charges under $\zk\times\overline{\bf Z}_k$ symmetry,
respectively. These currents have also conjugate partners, but they are
related as follows, 
\bea \nn \label{3.12}
\psi_l^\dagger(z)&=& \psi_{k-l}(z) \\
\bar{\psi}_l^\dagger(\bz)&=& \bar{\psi}_{k-l}(\bz) 
\eea

The currents algebra for the left handed currents is,
\bea \label{3.15}\nn
\psi_l(z)\psi_{l'}(w) &=& c_{ll'} (z-w)^{-2ll'/k}\psi_{l+l'}(w) +\cdots 
\hspace{1cm} l+l' < k \\
\psi_l(z)\psi_{l'}^\dagger(w) &=& c_{ll'}
(z-w)^{-2l(k-l)/k}\psi_{l-l'}(w)
+\cdots 
\hspace{1cm} l < l' \\
\psi_l(z)\psi_l^\dagger(w) &=& (z-w)^{-2l(k-l)/k}\lgroup I +
\frac{2d_l}{c}(z-w)^2 T(w)+\cdots \rgroup \nn
\eea
where $c_{ll'}$ is a constant, $I$ the identity operator, $T(z)$ the
energy-momentum tensor, $d_l$ the conformal weights of $\psi_l$ (the same
as $\psi_l^\dagger$) and $c$ the central charge. The two latter's are
given as,
\bea \label{3.20}
d_l=\frac{l(k-l)}{k}, \h1 {\rm and} \h1 c= \frac{2(k-1)}{k+2}.
\eea  
The relation between the parafermions and the original $SU(2)$ currents
can be seen from the following realization of $SU(2)$ algebra
\cite{zamolodchikov},
\bea \label{3.22} \nn
J_{+} &=& \frac{\sqrt{k}}{\sqrt{2}} \psi_1 \e
^{i\frac{\sqrt{2}}{\sqrt{k}}h} \\ 
J_{-} &=& \frac{\sqrt{k}}{\sqrt{2}} \psi_1^{\dagger} \e
^{-i\frac{\sqrt{2}}{\sqrt{k}}h} \\ 
J_3 &=& \frac{\sqrt{k}}{\sqrt{2}} i \p h  \nn
\eea 
The $SU(2)/U(1)$ is reached by gauging $J_3$ and it roughly means
to drop $h$ in (\ref{3.22}). The remaining currents $\psi_1$ and
$\psi_1^{\dagger}$ have not a closed algebra, instead they generate the 
$2k$-dimensional algebra (\ref{3.15}), for finite $k$.
 
Now, let $V[{\rm A}_a] = \sum_{j,n}a_a^{jn}V[jn]$ be an arbitrary 
linear combination of vertex operators, $V[jn]$. Then the OPE of currents 
and the primaries is as follows \cite{zamolodchikov},
\bea \label{3.25}
\psi_l(z)V[{\rm A}_a](w) = \frac{1}{(z-w)^{d_l}} V[{\rm T}_l{\rm
A}_a](w),
\eea
where ${\rm T}_l$ is the representation of $\psi_l$ in the given basis. 
The OPE of these vertex operators can be written as,
\bea \label{3.27}
V[jn](z) V[j'n'](w) = \sum_{j^{\"}n^{\"}}
(z-w)^{\Delta_{j^{\"} n^{\"}}-\Delta_{jn}-\Delta_{j'n'}}
V[j^{\"}n^{\"}](w),
\eea
in which $\Delta_{jn}= \frac{j(j+1)}{k+2}-\frac{n^2}{4k}$ is the conformal
weight of primaries and $[j^{\"}n^{\"}]$'s are all representations in the
fusion $[jn] * [j'n']$. 

Since we are looking for the low energy effective action, it is
sufficient to look at the decoupling limit \cite{SW} which corresponds to
massless states and can be reached by $k \goes \infty$. In this limit, it
can be seen from (\ref{3.15}) that $\psi_1$ and $\psi_1^{\dagger}$ have
a closed algebra in the following form:
\bea \label{3.30}
\psi_1(z)\psi_1^\dagger(w) = \frac{I}{(z-w)^2}
\eea
It is convenient to introduce ${\rm j}_1= (\psi_1 +
\psi_1^\dagger)/\sqrt{2}$  
and ${\rm j}_2= i(\psi_1 - \psi_1^\dagger)/\sqrt{2}$, then we have,
\bea \label{3.32}
{\rm j}_a(z) {\rm j}_b(w) = \frac{\delta_{ab}}{(z-w)^2}
\eea
\bea \label{3.33}
{\rm j}_a(z)V[{\rm A}_b](w) = \frac{1}{(z-w)} V[{\rm L}_a{\rm
A}_b](w),
\eea
where $a, b=1, 2$ and ${\rm L}_a$ is the representation of ${\rm j}_a$. 

Also the weights of vertex operators are zero for  $k \goes \infty$, so
from (\ref{3.27}), 
\bea \label{3.35}
V[{\rm A}_a](z) V[{\rm A}_b](w) = V[{\rm A}_a * {\rm A}_b](w),
\eea
where we have used $*$ as a notation for fusions of the representations.
However, this $*$ product indeed defines a non-commutative product. On the
flat
space the $*$ product comes from the non-commutativity of
coordinates on the boundary. For $SU(2)$ case, it can be found directly 
from the fusion algebra of $SU_k(2)$ (see \cite{ARS}). For the parafermionic
theory, it can be understood as a reduced form of $*$ product in $SU_k(2)$
theory. However, we postpone the definition to section 5.

With (\ref{3.32}) to (\ref{3.35}), the OPE's are similar to a free theory
in flat space \cite{ARS,SW} and it
is easy to find the 3-point function (\ref{3.10}). Also a similar
calculation can be done for 4-point function 
and the results are as follow, (see \cite{ARS} for details)
\bea \label{3.40}
S_{(3)} &\sim& {\rm L}_a{\rm A}_b * [{\rm A}^a , {\rm A}^b]_* \\
S_{(4)} &\sim& [{\rm A}_a , {\rm A}_b]_* * [{\rm A}^a , {\rm A}^b]_*
\eea

The quadratic or mass term of the action can be read from the mass
operator,
\bea \label{3.45}
M^2 = \frac{J_\mu J^\mu}{k+2}-\frac{(J_3)^2}{k}
\eea
In large $k$ limit it will be,
\bea \label{3.50}
M^2 = \frac{1}{k}\left(J_\mu J^\mu - (J_3)^2 \right) =
\frac{1}{k}(J_+J_- + J_- J_+)
\eea
The last expression is equivalent to $\psi_1\psi_1^{\dagger} \sim {\rm
j}_a {\rm j}^a \sim {\rm L}_a{\rm L}^a$, so we will have the following
quadratic term in the effective action,
\bea \label{3.55}
S_{(2)} \sim {\rm L}_a{\rm A}_b * {\rm L}_a{\rm A}_b
\eea

Now one can arrange the above terms in the following action,
\bea \label{3.60}
S = \frac{1}{4} \tr (F_{ab} * F^{ab})
\eea
where $F_{ab}= {\rm L}_a{\rm A}_b - {\rm L}_b{\rm A}_a + [{\rm A}^a , {\rm
A}^b]_*$. This result can be obtained by reduction from $SU(2)$ case as
has been done in \cite{schomtalk}.

Two remarks on this effective action are worth mentioning: 

\noindent 
$\bullet$ The action (\ref{3.60}) is equivalent to that on a
plane and there is no signature for the disk. Indeed, by sending $k \goes
\infty$, the disk goes to the entire plane. One can rescale coordinates on
the disk as $r \goes \tilde{r}/\sqrt{k}$, then the metric will be,
\bea \label{3.65}
ds^2 &=& \frac{k}{k-\tilde{r}^2}(d\tilde{r}^2 + \tilde{r}^2 d\phi^2)  
\eea
In limit $k \goes \infty$ with fixed $\tilde{r}$, this will be a flat
plane.

\noindent
$\bullet$ The $*$ product has not yet been identified with a geometric
construction. On the flat
space the $*$ product in the action comes from the non-commutativity of
coordinates on the boundary. For non-flat backgrounds such as the disk
(\ref{3.65}), with finite radius, we discuss a possible definition in
section 5.

\Section{Finite ${\bf k}$ Corrections to the Effective Action and
Fedosov's $\star$-product}

As mentioned in the end of the previous section, for $k \goes \infty$
limit, the disk is equivalent to a plane. On the other hands, a geometric
interpretation needs large $k$ limit, thus, to find a finite radius disk,
we have to consider $\frac{1}{k}$ corrections.  

Such corrections with a signal for a disk solution was found in
\cite{sahakian} in a different context. In \cite{sahakian}, the static
action is suggested to be,
\bea \label{4.10}
S = \int \tr \left( -\frac{1}{4} [X_a,X_b]^2 - \frac{1}{k} (X_a X^a)
\right).
\eea
where $X_a$ stands for $A_a$ in the previous section and can be understood
as coordinates of the background geometry. The second term in (\ref{4.10}) 
comes from a constant dilaton background or can be considered as
the leading order term of dilaton in $\sqrt{1-r^2}$. 

The equation of motion will be as follows,
\bea \label{4.15}
[X^a,[X_a,X_b]] - \frac{2}{k} X_b =0.
\eea
There are two interesting classes of solutions. Firstly, consider a
3-dimensional action ($a,b=1,2,3$). It has the following  
solution\cite{sahakian},
\bea \label{4.20}
X_a = \frac{1}{2\sqrt{k}} \tau_a, 
\eea 
where $\tau_a$'s are some irreducible representation of the $SU(2)$ Pauli
matrices. So we have,
\bea \label{4.25}
X_1^2 +X_2^2+X_3^2 = R^2
\eea
in which $R$ is a constant depending on $k$ and the dimension of 
irreducible representation.
The solution (\ref{4.20}) with (\ref{4.25}) is a 2-sphere.

Another class of solution to (\ref{4.15}) is the following, 
\bea \label{4.30}
X_1 = \frac{1}{\sqrt{2k}} \tau_1, \;\; X_2 = \frac{1}{\sqrt{2k}} \tau_2,
\;\; X_3=0.
\eea
This solution can be interpreted as a disk which is a squeezed version of
the sphere in (\ref{4.25}). 

Now we extend the action (\ref{4.10}) by adding next to leading order in
$1/k$ expansion of dilaton potential $\sqrt{1-r^2/k}=1- r^2/2k +\cdots$,
\bea \label{4.35}
S = \int \tr \left( - [X_1,X_2]^2 - \frac{1}{k} (X_1^2 +X_2^2) +
\frac{1}{2k^2} (X_1^2 +X_2^2)^2
\right).
\eea
with the following equation of motion,
\bea \label{4.40}
[X_1,[X_1,X_2]] - \frac{2}{k} X_2(1-\frac{r^2}{k}) =0.
\eea
In (\ref{4.35}) and (\ref{4.40}), we have replaced $X_a \goes
X_a/\sqrt{k}$ to have correct large $k$ limit. The equation of motion
(\ref{4.40}) has a solution as,
\bea \label{4.45}
[X_1,X_2] =-i(1-\frac{r^2}{k}) 
\eea

To interpret this solution as a disk, we need to define the products in
the above relations correctly. Indeed, (\ref{4.45}) can define a $\star$
product as follows,
\bea \label{4.50}
X_1 \star X_2 - X_2 \star X_1 = -i(1-\frac{X_1^2 + X_2^2}{k}).
\eea
In $k \goes \infty$ limit, the commutator (\ref{4.45}) is $[X_1, X_2]=-i$
which is a plane brane solution in flat space.

It can be shown that the $\star$ product (\ref{4.50}) is consistent with a
natural non-commutative structure on the given disk in (\ref{1.35}). This
non-commutative structure is obtained from a deformation quantization
procedure known as Fedosov's formalism. In this formalism, the basic idea
is to find a map between the space of functions on a symplectic curved
manifold to its Weyl algebra bundle on which the ordinary Moyal product
can be defined. By inverting the map, one can find the product rule for
functions on the manifold.

Here, we will briefly give the basic prescription based on 
\cite{kishimoto} for circularly symmetric two dimensional spaces. 
Consider a symplectic manifold $M$ with $\Omega_0$ as its symplectic form.
Let $W$ be the Weyl algebra bundle with an Abelian connection $D$ and an
$\circ$ product as the ordinary Moyal product:
\bea \label{4.55}
\circ := \exp(\frac{i\hbar}{2}\frac{\overleftarrow{\p}}{\p y^i}
\omega^{ij} \frac{\overrightarrow{\p}}{\p y^j})
\eea
where $\omega^{ij}$ is a constant antisymmetric parameter and $\hbar$ is
a deformation parameter. Now for $W_D\equiv ker D \subset W$, we can
define an invertible map from $C^{\infty}(M)[[\hbar]]$ to $W_D$, 
\bea \label{4.60}
Q \; : \;\;\; C^{\infty}(M)[[\hbar]] \goes W_D  
\eea
with inverse:
\bea \label{4.65}
\sigma \; : \;\;\; W_D \goes C^{\infty}(M)[[\hbar]] 
\eea
Then Fedosov's $\star$ product on  $C^{\infty}(M)[[\hbar]]$ can be defined
by,
\bea \label{4.70}
a_{0}\star b_{0}:= \sigma \left(Q(a_0) \circ Q(b_o)\right),  \;\;\;
a_0, b_0 \in C^{\infty}(M)[[\hbar]]
\eea
This product is associative and is a deformation of the Poisson bracket
with the symplectic form $\Omega_0$.

Now for a two dimensional circularly symmetric space,
\bea \label{4.75}
ds^2 = \Lambda (r) ( dr^2 +r^2 d\phi^2),
\eea
the map $Q$ can be defined as follows \cite{kishimoto},
\bea \label{4.80}
a = Q\left( a_0(r,\phi)\right)= a_0 \left( G(r,y_1), \phi
+\frac{y_2}{r}\right) \\ \nn
a \in W_D, \;\;\;  a_0 \in  C^{\infty}(M)[[\hbar]]
\eea
where $G(r,y_1)$ is found in,
\bea \label{4.85}
\int^{G(r,y_1)}_r \Lambda (r') dr' = y_1 r
\eea
With $Q$ in hand, one can define the $\star$ product on $M$ by
(\ref{4.70}),
\bea \label{4.87}
a(r,\phi) \star b(r,\phi) =  \left(a_0 \left( G(r,y_1), \phi
+\frac{y_2}{r}\right) 
\circ b_0 \left( G(r,y_1), \phi +\frac{y_2}{r}\right) \right)_{y_1=y_2=0}
\eea

Now consider the special case of the disk metric in (\ref{3.65}) for which
$\Lambda (r) = k/(k-r^2)$ and $G(r,y_1)$ can be found as,
\bea \label{4.90}
G(r,y_1)=\sqrt{k-(k-r^2)\e^{-2y_1r/k}}
\eea
It is simple to compute the $\star$ product for $X_1=r \cos \phi$ and
$X_2=r \sin \phi$ as functions on $M$,
\bea \label{4.95}
X_1 \star X_2 - X_2 \star X_1 = -i\hbar(1-\frac{r^2}{k}).
\eea
It is just equivalent to the relation (\ref{4.45}) (with $\hbar=1$). This
means that Fedosov's formalism on the disk confirms the action
(\ref{4.35}) with the solution (\ref{4.45}).

\section{Conclusion}

The effective action for branes on $SU_k(2)/U_k(1)$ gauged WZW is found
to be a non-commutative gauge theory on a disk, for infinite $k$ limit.
This effective action is equivalent to that for a plane, and this is not
surprising, since the disk is a plane for $k \goes \infty$.

On the other hand, finite $k$ corrections are possible, 
which correspond to new terms in the matrix model/DBI, respecting
non-constant dilaton background. 
The resulting action with $1/k$
corrections is a non-commutative theory with a special kind of
non-commutative algebra. We have shown that the non-commutativity can be
understood in the Fedosov's deformation quantization on the disk. 

\section*{Acknowledgement}

The author is very grateful to G. Mandal and S. Wadia for bringing his
attention to this problem along with very useful discussions and
correspondence. Thanks also to A. Dabholkar, S. Mukhi, and N. V.
Suryanarayana for comments and also S. Trivedi for discussion. This work
was partially supported by a travel grant from the Institute for Studies
in Theoretical Physics and Mathematics, Tehran, Iran.

%




\begin{thebibliography}{99}

\bibitem{polchinski}  J. Polchinski, {\it ``String Theory"},
Cambridge Univ. Press, 1998.

\bibitem{green}  M.B. Green and P. Wai, Nucl. Phys. {\bf B431} (1994)
131; M.B. Green, Nucl. Phys. {\bf B381} (1992) 201; M.B. Green  and
M. Gutperle,  Nucl. Phys. {\bf B476} (1996) 484, hep-th/9604091.

\bibitem{ishibashi} N. Ishibashi and T. Onogi, Nucl. Phys. {\bf B318}
(1989) 239;

\bibitem{cardy} J.L. Cardy,  Nucl. Phys. {\bf B324} (1989) 581.

\bibitem{NC} M.R. Douglas and C. Hull, JHEP {\bf 9802}
(1998) 008, hep-th/9711165; \\
Y.-K.E. Cheung and M. Krogh, Nucl. Phys. {\bf B528} (1998) 185,
hep-th/9803031; \\
F. Ardalan, H. Arfaei and M.M. Sheikh-Jabbari, JHEP {\bf
9902} (1999) 016, hep-th/9810072; \\
H. Garcia-Compe$\acute{\rm a}$n,  Nucl. Phys. {\bf B541} (1999) 651
hep-th/9804188; 

\bibitem{SW} N. Seiberg and E. Witten, JHEP {\bf 9909} (1999) 032,  
hep-th/9908142.

\bibitem{dgroup} M. Kato and T. Okada, Nucl. Phys. {\bf B499} (1997) 583,
hep-th/9612148; \\
A. Recknagel and V. Schomerus, Nucl. Phys. {\bf B531}
(1998) 185, hep-th/9712186;\\
A. Alekseev and V. Schomerus, Phys. Rev. {\bf D60} (1999)
061901, hep-th/9812193.

\bibitem{ARS}  A. Alekseev, A. Recknagel and V. Schomerus, 
JHEP {\bf 9909} (1999) 023, hep-th/9908040; 
JHEP {\bf 0005} (2000) 010, hep-th/0003187; 
Mod. Phys. Lett. {\bf A16} (2001) 325, hep-th/0104054.

\bibitem{mald}  J. Maldacena, G. Moore and N. Seiberg, {\it ``Geometrical
interpretation of D-branes in gauged WZW models"}, hep-th/0105038.

\bibitem{zamolodchikov}  A.B. Zamolodchikov and V.A. Fateev, Sov. Phys.
JETP {\bf 62} (1985) 215;\\
D. Gepner and Z. Qiu, Nucl. Phys. {\bf B285} (1987) 423.

\bibitem{bachas} C. Bachas, M. Douglas and C. Schweigert, JHEP {\bf 0005}
(2000) 048, hep-th/0003037.

\bibitem{schomtalk} S. Fredenhagen and V. Schomerus, {\it ``Brane Dynamics
in CFT Backgrounds"}, Talk given at Strings 2001, 5-10 Jan 2001, Mumbai,
India,  hep-th/0104043.

\bibitem{sahakian} V. Sahakian, JHEP {\bf 0104} (2001) 038, 
hep-th/0102200.

\bibitem{Fedosov} B.V. Fedosov, {\it ``Deformation quantization and index
theory"}, Berlin, Germany, Akademie-Verl. (1996);\\
T. Askawa and I. Kishimoto, Nucl. Phys. {\bf B591} (2000) 611,
hep-th/0002138.

\bibitem{kishimoto} I. Kishimoto, JHEP {\bf 0103} (2001) 025,
hep-th/0103018.

\end{thebibliography}
\end{document}